\documentclass[twocolumn,jap,showpacs,floatfix, ]{revtex4}  
\usepackage{color,graphicx}
\usepackage{verbatim}
\usepackage{amssymb}
\usepackage[normalem]{ulem}
\usepackage{latexsym}

\definecolor{red}{rgb}{0.8,0,0.2}

\def\beeq{\begin{equation}}
\def\eneq{\end{equation}}
\def\beeqa{\begin{eqnarray}}
\def\eneqa{\end{eqnarray}}

\begin{document}
\DeclareGraphicsExtensions{.ps,.pdf,.eps, .jpg}

 \title{Thermally-activated charge reversibility of gallium vacancies in GaAs}
 
\author{Fedwa El-Mellouhi} \email {f.el.mellouhi@umontreal.ca}
 \affiliation{D\'epartement de physique and Regroupement qu\'eb\'ecois
   sur les mat\'eriaux de pointe, Universit\'e de Montr\'eal, C.P.
   6128, succ. Centre-ville, Montr\'eal (Qu\'ebec) H3C 3J7, Canada}

\author{Normand Mousseau } \email {Normand.Mousseau@umontreal.ca}
 \affiliation{D\'epartement de physique and Regroupement qu\'eb\'ecois
   sur les mat\'eriaux de pointe, Universit\'e de Montr\'eal, C.P.
   6128, succ. Centre-ville, Montr\'eal (Qu\'ebec) H3C 3J7,
   Canada\footnote{permanent address}}

\affiliation{Service de Recherches de M\'etallurgie Physique, Commissariat  \`a
  l'\'energie atomique-Saclay, 91191 Gif-sur-Yvette, France.}

\date{\today}

\begin{abstract}
  The dominant charge state for the Ga vacancy in GaAs has been the subject of a
long debate, with experiments proposing $-$1, $-$2 or $-$3 as the best answer.
We revisit this problem using {\it ab initio} calculations to compute the
effects of temperature on the Gibbs free energy of formation, and we
find that the thermal dependence of the Fermi level and of the ionization
levels lead to a reversal of the preferred charge state as the temperature
increases. Calculating the concentrations of gallium vacancies based on these
results, we reproduce two conflicting experimental measurements, showing that
these can be understood from a single set of coherent LDA results when thermal
effects are included.
\end{abstract}


\maketitle
\section{Introduction}

There has been a long-standing controversy about the nature of the charge
state of the gallium vacancy ($V_{Ga}^q$) mediating self-diffusion in GaAs.
While experiments have deduced a preferred charge state ranging from $-$1 and
$-$2~\cite{4Bra99_2} to $-$3~\cite{4Geb03, 4Tan93} under N-type or intrinsic
conditions, most (zero temperature) {\it ab-initio} calculations favor the
$-$3 charge state for $V_{Ga}$, see Ref.~\cite{4Elm05,4Zha91, 4Laa92, 4Che94,4Seo95,4Nor93, 4Poy96,
4Sch02, 4Jan03, 4Jan89, 4Bar85}.
Extracting the charge state from experiments is challenging: these are usually
performed at high temperature, in a regime where the formation entropy becomes
a significant part of the Gibbs free energy. These entropic contributions can
therefore alter the charge of the dominant species as calculated at $T=0$,
leading to a strong competition between the charge states. Moreover, the
separate ionization energy for each state is not directly accessible
experimentally; only the sum over the ionization free energies can be obtained
by fitting the experimental data over a range of temperatures, using a number
of assumptions typically inspired from {\it ab-initio} data~\cite{4Geb03,
4Tan93} or set as free parameters in the fit~\cite{4Bra99_2}.

Since the ionization free energies are key for fitting experimental data, it
is important to evaluate them from {\it ab-initio} calculations. Because of
the high computational cost associated with computing vibrational
spectra~\cite{4Raul04}, however, various assumptions are made typically to
simplify the problem. For example, it is usually assumed that the free energy
of ionization is independent of temperature and that ionization levels are
constant with respect to the valence band edge so that the formation entropy
of all charge defects is taken as equal to that of $V_{Ga}^0$~\cite{4Boc96}.
Often, it is also supposed that ionization energies follow the thermal
behavior of band gap (or the conduction band minimum)~\cite{4Van76}, which
results typically into shallower ionization levels with increasing temperature
(within this approximation, for example, heavily charged defects in Si become
favored at high T~\cite{4Dev03}).

In this article, we revisit the question of the dominant charge state of
$V_{Ga}$ in GaAs, calculating directly, using LDA, the ionization Gibbs free
energies ($F^{q1/q2}$) for all possible charge state transitions, using an
approach applied to SiC~\cite{4Raul04} and Si~\cite{4San05}. We show that, in
spite of well known LDA limitations, apparently conflicting data coming from
two experiments performed under different conditions~\cite{4Bra99_2, 4Geb03} are
well fitted. This satisfactory fit is obtained for both experimental data by
using the same set of inputs like the concentration model, the Fermi level
effect and the computed temperature dependence of $F^{q1/q2}$.

This article is organized as follows: calculation details are presented in
section~\ref{sec:comp}, next we analyze the source of conflict that arose
between different experiments in section~\ref{sec:anal}, presenting at the
same time the common concentration and diffusion models used.
Section~\ref{sec:res} contains a discussion of phonon calculation results
followed by our fitting procedure and its application on two experimental
data. The main conclusions are presented in section~\ref{sec:conclusion}.

\section{Computational details}
\label{sec:comp} 

From the computational point of view, the stability of $V_{Ga}^q$ is
determined by comparing total energies of supercells with and without defects.
In a previous ground state {\it ab-initio} calculation using SIESTA, a
local-basis LDA code~\cite{4Elm05}, we computed the ionization levels of
gallium vacancies for  0, $-$1, $-$2, $-$3 charges.

Here, we include the thermal contributions by computing the Gibbs free
energies in the harmonic limit. Note that entropy changes in bulk  Ga and As metals is accounted for (see Refs.~\cite{4Raul03, 4Raul04} as well as
Ref.~\cite{4Elm05} for more details):
 \begin{equation}
 G = E_{tot} + U_{vib} - TS_{vib} + p.V + \sum_i \mu_i N_i
 \end{equation}
where, at constant number of particles and volume, the vibrational energy and entropy are given, respectively, by
\begin{equation}
U_{vib} = \sum_{i=1} ^{3N} \left\{ {{\hbar \omega_i} \over{\exp(\hbar
\omega_i/k_B T)-1}} +{ 1\over2} \hbar \omega_i \right\}
\end{equation}
and
\begin{equation}
  S_{vib} = k_B \sum_{i=1} ^{3N} \{{ {\hbar \omega_i} \over {k_B T}}
    [\exp ({{\hbar \omega_i}\over {k_B T}}) -1]^{-1} - 
    \ln [1-\exp({{-\hbar
            \omega_i}\over {k_B T}}) ]\}.
\end{equation}
To obtain the Gibbs free energy, the phonon spectrum ($\omega_i$) is computed
for all stable charge states of $V_{Ga}^q$ using a numerical derivative to
construct the dynamical matrix.

The calculation of the dynamical matrix of cells with 215 sites is a very
time-consuming task. Since many atomic shells around the vacancy are distorted
it is not possible to reduce the number of calculations by symmetry
operations. We thus use the full dynamical matrix for the complete
supercell. A mesh cutoff of 120 Ry is used and the Brillouin zone of the
215-atom cubic supercell is sampled at the $\Gamma$-point. The eigen
frequencies are then extracted at 146 k-points in the Brillouin
zone~\cite{4San05}.

Except for the splitting of the optical branches at the $\Gamma$-point, which
cannot be obtained in the harmonic approximation, the total phonon density of
states and the dispersion curve obtained by this procedure for the GaAs
crystal are in good agreement with experiment and calculations~\cite{4Pet97,
4Str90}. Since we are interested here in the entropy difference and not
its absolute value, these discrepancies should not play an important role.

\section{Analysis of the experimental results}
\label{sec:anal}

Before discussing our analysis, we first focus on the source of discrepancy
that exists in explaining different experiments. Although the diffusion model
and the experimental Fermi level dependence used in different
experiments~\cite{4Bra99_2, 4Geb03} are the same, conclusions about the dominant
charge state of $V_{Ga}$ are contradictory. This contradiction can therefore
be due to experimental errors or the additional assumptions about the
temperature dependence of the ionization levels used in the experimental fits.
The following analysis shows that the choice of the ionization levels is an
important factor that suffices to explain the conflict. Some calculation
details are skipped in the manuscript, we refer the reader to Refs.\
~\cite{4Bra99_2,4Tan93} for further details on the model used.
 
\subsection{Concentration model}

Both fitting procedures used by Gebauer {\it et al.}~\cite{4Geb03} and Bracht
{\it et al.}~\cite{4Bra99_2} are based on the concentration model used previously
by Tan and coworkers~\cite{4Tan93}. The Ga self-diffusion coefficient $D_{Ga}$
is given by the sum of the transport coefficients of vacancies in various
charge states $q$:
\beeq
\label{eq:bra1}
D_{Ga} = {1 \over C_0}\sum_{q=0}^3f_qC^{eq}_{V_{Ga}^{-q}}D_{V_{Ga}^{-q}}
\eneq
where, $ C_0$ represents the Ga atom density in GaAs
($C_0=2.215\times10^{22}\text{cm}^{-1}$), $f_q$ is the diffusion correlation factor
that contains information about the microscopic jump mechanism,
$C^{eq}_{V_{Ga}^{-q}}$ is the thermal equilibrium concentration and
$D_{V_{Ga}^{-q}}$ is the diffusion coefficient of the vacancy $V_{Ga}^{-q}$
for $q ={0,1,2,3}$.

The thermal equilibrium concentration of $V_{Ga}^{-q}$ in GaAs is defined as
\beeq 
\label{eq:bra2} 
C^{eq}_{V_{Ga}^{-q}} = C^* P_{As_4}^{1/4}
T^{-5/8}\exp\left(\displaystyle{-\frac{G^f_{V_{Ga}^{-q}}}{k_BT}}\right),
\eneq
where $C^*$ is a pre-exponential factor, $G^{f}_{V_{Ga}^{-q}}$ is the Gibbs
free energy of formation of $V_{Ga}^{-q}$, $T$ is temperature and $k_B$ and is
the Boltzmann constant. This equation takes into account the influence of
$As_4$ over-pressure ($ P_{As_4}$), often used in experiments, on
vacancy concentrations. If the arsenic over-pressure increases, it affects the
stoichiometry  of gallium arsenide by introducing an excess of As atoms. With
this excess of As atoms, GaAs sample can become As-rich and the concentrations
of Ga vacancies in GaAs are enhanced.

Different vacancies can introduce energy levels within the energy
band gap of GaAs. The occupation of these energy states depends on the
position of the Fermi level. Under extrinsic conditions, when the hole or the
electron concentration introduced by doping exceeds the intrinsic carrier
concentration, the Fermi level deviates from its intrinsic position. As a
consequence, the ratio of the charged to neutral vacancy concentrations is
changed. Vacancies can introduce acceptor levels, i.e., $V_{Ga}^{-1}$
introduces one level located at $E_{V_{Ga}^{-1}}$ above the valence band edge,
$V_{Ga}^{-2}$ introduces  one at
$E_{V_{Ga}^{-2}}$ and $V_{Ga}^{-3}$ introduces 
one at $E_{V_{Ga}^{-3}}$ above the valence band edge.

The concentration profiles measured experimentally contain  contributions from  all active  vacancy charge states that cannot be separated.  We take advantage of the fact that the Gibbs free energy of formation --- and thus the equilibrium concentration    $C^{eq}_{V_{Ga}^{0}}$ --- of neutral vacancies do not depend on the position of the extrinsic Fermi level. Therefore, we can express the  total concentration  as function of the ratio of
the charged to neutral vacancy concentrations:

\beeq
\label{eq:bra6}
C^{eq}_{V_{Ga}} = \displaystyle \sum^3_{q=0} C^{eq}_{V_{Ga}^{-q}}=C^{eq}_{V_{Ga}^{0}}\sum^3_{q=0} {\frac{C^{eq}_{V_{Ga}^{-q}}}{C^{eq}_{V_{Ga}^{0}}}}
\eneq
The ratios are given by:
\beeqa
\label{eq:bra3}
{\frac{C^{eq}_{V_{Ga}^{-}}}{C^{eq}_{V_{Ga}^{0}}}} &=& g_{V_{Ga}^{-}}\exp\left(\displaystyle{\frac{E_f-E_{V_{Ga}^{-}}}{k_BT}}\right)\\
\label{eq:bra4}
{\frac{C^{eq}_{V_{Ga}^{-2}}}{C^{eq}_{V_{Ga}^{0}}}} &=& g_{V_{Ga}^{-2}}\exp\left(\displaystyle{\frac{2E_f-E_{V_{Ga}^{-2}}-E_{V_{Ga}^{-}}}{k_BT}}\right)\\
\label{eq:bra5}
{\frac{C^{eq}_{V_{Ga}^{-3}}} {C^{eq}_{V_{Ga}^{0}}}} &=& g_{V_{Ga}^{-3}}\exp\left(\displaystyle{\frac{3E_f-E_{V_{Ga}^{-3}}- E_{V_{Ga}^{-2}}-E_{V_{Ga}^{-}}}{k_BT}}\right)
\eneqa
where $g_{V_{Ga}^{-q}}$ is approximated to unity (further explanations and intermediate steps used in this work can be found in Ref.~\cite{4Bra99_2}).

The concentration of negatively charged vacancies can be deduced if we know
the position of the Fermi level $E_f$ and the ionization energies
$E_{V_{Ga}^{-q}}$. Ionization levels do not cause much problems
as they are derived from available {\it ab-initio} calculation or
included in the fit. However, the Fermi level depends strongly on intrinsic
($n_i$) and extrinsic carrier concentrations ($n$) by the relation:
\beeq
\label{eq:bra8}
{n \over {n_i}} = \exp\left(\displaystyle{\frac{E_f-E_f^i}{k_BT}}\right)
\eneq
Intrinsic properties of GaAs as function of temperature are taken from
empirical data~\cite{4Bla82}, like intrinsic band gap $E_G$, electron and hole
concentrations ($n_i$ and $p_i$) and intrinsic Fermi level $E_f^i$. The band
gap of GaAs is supposed to vary with temperature~\cite{4Thu75}, following
Varshni's~\cite{4Var67} empirical expression for GaAs $E_G(T)~=~1.519 -
5.0405*10^{-4}T^2/(T + 204)$ deduced  from optical experiments~\cite{4Bla82}. The Fermi energy is
the important quantity  governing the equilibrium
concentrations of defects. However, it changes if any electrically active
impurities or dopants are present in the material as well as under the
presence of charged vacancies. The concentrations of these impurities or
dopants ($C_{Donor}, C_{Acceptor}$) and vacancy concentration in all charge
states ($C^{eq}_{V_{Ga}^{-q}}$) must therefore be taken into account.
 The charge balance equation, requiring that the free
electron and hole concentrations $n$ and $p$ (which also depend on the Fermi
level) must cancel out any net charge resulting from the concentrations of all
positively and negatively charged defects and impurities:
\beeqa
\label{eq:bra9} 
n&=& {1\over 2}\left(C_{Donor} - \displaystyle
\sum_{m=0}^3 mC^{eq}_{V_{Ga}^{-q}}\right) \nonumber \\ &&
+ \sqrt{n_i^2 +
{1\over4}\left(C_{Donor} -\displaystyle \sum_{m=0}^3
mC^{eq}_{V_{Ga}^{-q}}\right)^2} 
\eneqa
\beeqa
\label{eq:bra10}
{{n_i^2}\over n}&=&p= {1\over 2}\left(C_{Acceptor} + \displaystyle \sum_{m=0}^3 mC^{eq}_{V_{Ga}^{-q}}\right) \nonumber \\ &&
+ \sqrt{n_i^2 + {1\over4}\left(C_{Acceptor} +\displaystyle \sum_{m=0}^3 mC^{eq}_{V_{Ga}^{-q}}\right)^2}
\eneqa

The extrinsic Fermi level, for its part, is obtained by solving
simultaneously Equations \ref{eq:bra8}, \ref{eq:bra9}, \ref{eq:bra3},
\ref{eq:bra4} and \ref{eq:bra5} self-consistently using the Newton Raphson
method~\cite{4NR}. Once the Fermi level is known, it is used to determine the
final Gibbs free energy of formation and the corresponding equilibrium
concentrations of all the defects present in the structure. Up to this point,
the procedure is sufficient to reproduce the fit of Gebauer {\it et al.} and
Bracht {\it et al.} shown in Fig. 7 of Ref.~\cite{4Geb03} and Fig. 9
of Ref.~\cite{4Bra99_2}, respectively.

\subsection{Diffusion model}

Few steps are left to fit the measured diffusion coefficients of Bracht {\it
et al.} (see the detailed discussion in Section 5.2 of Ref.~\cite{4Bra99_2} ).
Based on the theoretical work of Bockstedte and Scheffler~\cite{4Boc96}, they
assume that the entropy change associated with the $V_{Ga}^{-q}$ migration and
the correlation factor $f_q$ are similar for all charge states, i.e.,
${D_{V_{Ga}^q}}\approx{ D_{V_{Ga}}}$ and ( $f_q\approx f=1$). In this approximation, Eq.~\ref{eq:bra1} simplifies to:
\beeq
\label{eq:bra12}
D_{Ga} = {1\over{C_0}}C^{eq}_{V_{Ga}^0} D_{V_{Ga}} {\displaystyle \sum_{q=0}^3{\frac{C^{eq}_{V_{Ga}^{-q}}}{C^{eq}_{V_{Ga}^0}}}}
\eneq
While the extrinsic diffusion coefficient ($D_{V_{Ga}}$) is unknown, it is, however, related to the intrinsic diffusion coefficient  by the relation:
\beeq
\label{eq:bra13}
{\frac{1}{C_0}}C^{eq}_{V_{Ga}^0} D_{V_{Ga}^0}=
\displaystyle{\frac{D_{Ga}(n_i)}{1+\displaystyle\sum_{m=1}^3 \exp{\frac{mE_f^i -\displaystyle\sum_{m=1}^3 E_{V_{Ga}^{-q}}}{k_BT}}}}
\eneq
The missing quantity in this equation is $D_{Ga}(n_i)$. Bracht {\it et al.}
use the Ga diffusion coefficient they measured in {\bf intrinsic} GaAs and
find $D_{Ga}$, using Eqns~\ref{eq:bra12}, \ref{eq:bra13}  and~:
 \beeq
 \label{eq:bra14}
 D_{Ga} (n_i)= 0.64\exp\left( {{-3.71}\over {k_BT}}\right)cm^2s^{-1}
 \eneq

\subsection{Fits of positron-annihilation measurements}

From positron-annihilation experiments, Gebauer {\it et al.}~\cite{4Geb03}
deduce the Gibbs free energy of formation and the dominant charge state of
$V_{Ga}$ by measuring directly vacancy concentrations. Measures were done at
different high temperatures (900-1450K) by changing dopants concentrations
and the chemical potential (stoichiometry).

The resulting concentration profiles were fitted using the model presented
above~\cite{4Tan93} using {\it ab-initio} ionization levels calculated by
Baraff et Schl\"uter~\cite{4Bar85} at 0K corresponding to $F^{0/-1}$~=~0.13$E_G$ , $F^{-1/-2}$~=~0.35$E_G$ and $F^{-2/-3}$~=~0.49$E_G$. These
ionization levels are supposed to vary with temperature like the band gap
following Varshni's~\cite{4Var67} empirical relation.

Gebauer {\it et al.}~\cite{4Geb03} fitted the concentration profiles for N-GaAs
with $[Te]~=~2\times 10^{18}\text{cm}^{-3}$ at arsenic over pressure of $P_{{As}_4}$~=~5.6~atm for temperature ranging between 900 and 1450K. The resulting fit
(corresponding to Fig. 7 of their article) is in reasonable agreement with
measured concentration profiles especially at high temperature. It shows a
negative temperature dependence that cannot be reproduced unless the $-$3
charge state dominates all other charge states. 

\begin{figure}[t]
\centerline{\includegraphics[width=7cm, angle=-90]{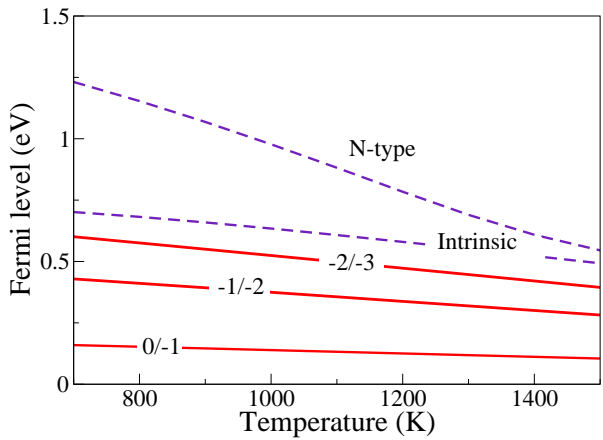}}
\caption[Ionization levels (solid lines) used by Gebauer {\it et
al.}~\cite{4Geb03}.]{Ionization levels (solid lines) used by Gebauer {\it et
al.}~\cite{4Geb03}, they are taken to have a negative
temperature dependence and vary like the experimental band gap. The position
of the extrinsic Fermi level is also shown (dash line). }
\label{fig:geb}
\end{figure}

Fig.~\ref{fig:geb}
shows the variation of ionization levels used as function of temperature, the
extrinsic Fermi level corresponding to the the dopant concentration used in
this experiment is also shown. We can see that this choice of the ionization
levels favors the stability of the $-3$ charge state for N-type and intrinsic
GaAs over a broad temperature range.

\subsection{Fits of interdiffusion measurements}

For their part, Bracht {\it et al.}~\cite{4Bra99_2}, measured interdiffusion in
isotope heterostructure $^{71}\text{GaAs}/^{\text{nat}}\text{GaAs}$ exposed to arsenic over
pressure of $P_{As}~=~1$ atm under intrinsic and extrinsic conditions (N-type
with $[Si]~=~3\times 10^{18}\text{cm}^{-3}$ and P-type with $[Be]~=~3\times
10^{18}\text{cm}^{-3}$). They measured the temperature dependence of the Ga
self-diffusion coefficient ($D_{Ga}$) at high temperature ranging between
1032K and 1350K.

\begin{figure}[t]
 \centerline{\includegraphics[width=7cm,angle=-90]{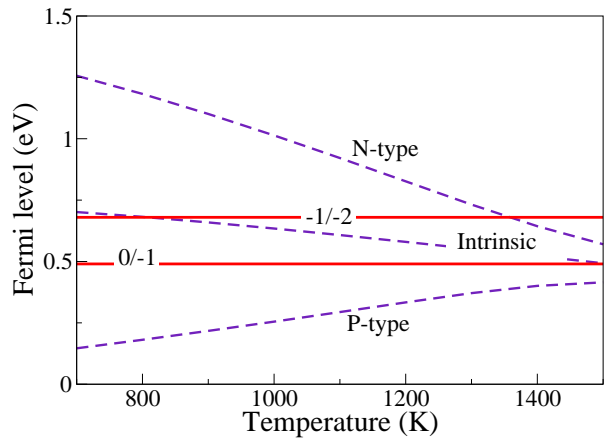}} %
\caption[Ionization levels (solid lines ) as obtained by Bracht {\it et
al.}~\cite{4Bra99_2}.]{Ionization levels (solid lines ) as obtained by Bracht {\it et
al.}~\cite{4Bra99_2} from a fit of the experimental data in the temperature
range 1032K-1350K and are assumed to remain constant. The position of the
extrinsic Fermi level is also shown (dash line). }

\label{fig:bra}
\end{figure}

The simultaneous fit of the three curves (N-type, P-type and intrinsic curves
in Fig. 10 of Ref.~\cite{4Bra99_2}) is complicated and lengthy,
especially that the ionization levels of Baraff and  Schl\"uter did not give a
satisfactory fit. Therefore they include the ionization levels in the fitting
procedure. The best fit obtained corresponds to $H_f~=~1.9$ eV and includes two
ionization levels only ($0/-1$ and $-1/-2$) located at $F^{0/-1}$~=~0.42~eV
and $F^{-1/-2}$~=~0.6~eV from the VBM that must remain unchanged with
temperature.

Figure~\ref{fig:bra} shows the variation of ionization levels used as function
of temperature. The Fermi level corresponding to intrinsic, N-type and P-type
GaAs is also shown. By analyzing this figure, one can notice that the $-2$
charge state dominates at low temperature, in intrinsic and N-type GaAs, then it 
is reversed to $-1$ and $0$ as temperature increases. In this temperature
range there is a competition between different charge states for intrinsic and
N-type GaAs with no contribution coming from the $-3$ charge state. This
choice of ionization level favors the stability of less charged vacancies as
temperature increases. For P-type GaAs, the neutral vacancy dominates
self-diffusion over the entire temperature range.

In both  experiments the dominant part of the
temperature dependence is due to the change of the chemical potential (i.e.,
the Fermi-level effect),  treated extensively by Tan and
coworkers~\cite{4Tan93}. However, the above analysis shows unequivocally that
only including the experimentally-measured thermal dependence of the Fermi
level alone is not sufficient to explain the two conflicting experiments at
the same time. Figures~\ref{fig:geb} and~\ref{fig:bra} show the
crucial role played by the ionization level behaviour for determining the
charge state of the vacancy and that they are the main source of conflict.

\section{Results and discussion}
\label{sec:res}

\begin{figure}[t]
\centerline{\rotatebox{-90}{\includegraphics[width=7cm]{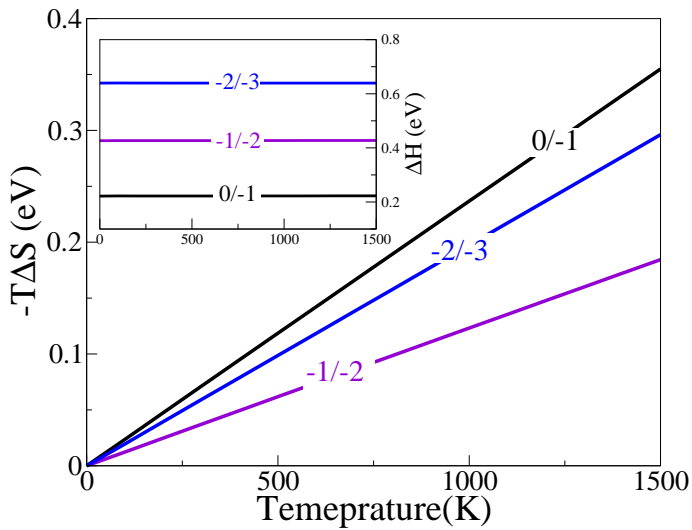}}}
\caption[Entropy and enthalpy (inset) differences obtained
after progressive ionization of $V^q_{Ga}$.]{ Entropy and enthalpy (inset) differences obtained
after progressive ionization of $V^q_{Ga}$. While $\Delta H^{q1/q2}$ shows
little temperature dependence, the non-vanishing entropy difference between
charge states lead to a linearly increasing entropic term, $T\Delta S^{q1/q2}~=~T(S^{q2} -S^{q1} )$, in the Gibbs free energy.}
\label{fig:HS}
\end{figure}

The calculated thermal effect on the enthalpy $\Delta H^{(q1/q2)}$ and the
entropy differences $T\Delta S^{(q1/q2)}$ between the relevant sequential
charge states are shown in Fig.~\ref{fig:HS}. For all differences, the
ionization enthalpy is independent of the temperature. In contrast, since the
entropy difference between these states is finite and mostly independent of T,
the ionization entropy term increases linearly with temperature. Moreover,
contrary to what could be expected, the neutral vacancy has the highest
entropy so that the formation entropy of $V_{Ga}$ should decrease with an
increase in the number of electrons. Important changes with temperature occur
in the band gap, and, depending on the doping type, in the Fermi level.
This temperature dependance of the Fermi level affects the
relative energy of neighboring charge states by the same amount. 
It is this effect that is responsible for the interesting effect of {\it
negative temperature dependance} of the $V_{Ga}$ concentration discussed in
Refs.~\cite{4Geb03, 4Tan93}.

While the change in the charge transition level due to entropy term is smaller
than the shift in the Fermi level --- the ($-$2/$-$3) transition is moved by
about 0.3 eV --- its effect is charge dependant, affecting the relation
between the various levels. At 0K, $V_{Ga}$ ionization levels in GaAs are
located in the lower part of the band gap and distant by 0.1$-$0.2 eV from
each other. Hence, an increase of 0.3 eV coupled with the Fermi level effect
cause important changes in the dominant vacancy charge.

Similar entropic effects are known to affect the properties of defects in
semiconductors. For example, the entropy term of amplitude similar to ours
appears to be responsible for the lowering the migration barrier in SiC with
increasing temperature~\cite{4Raul04}. It is also associated with the
destabilization of a metastable configuration in GaAs~\cite{4Ham88, 4Fan01} and
the thermal behaviour of the binding free energy of acceptor-oxygen complexes
in Si~\cite{4San05}.

\begin{figure}[t]
\centerline{\rotatebox{-90}{\includegraphics[viewport=150 40 400 500, angle=90, width=9cm]{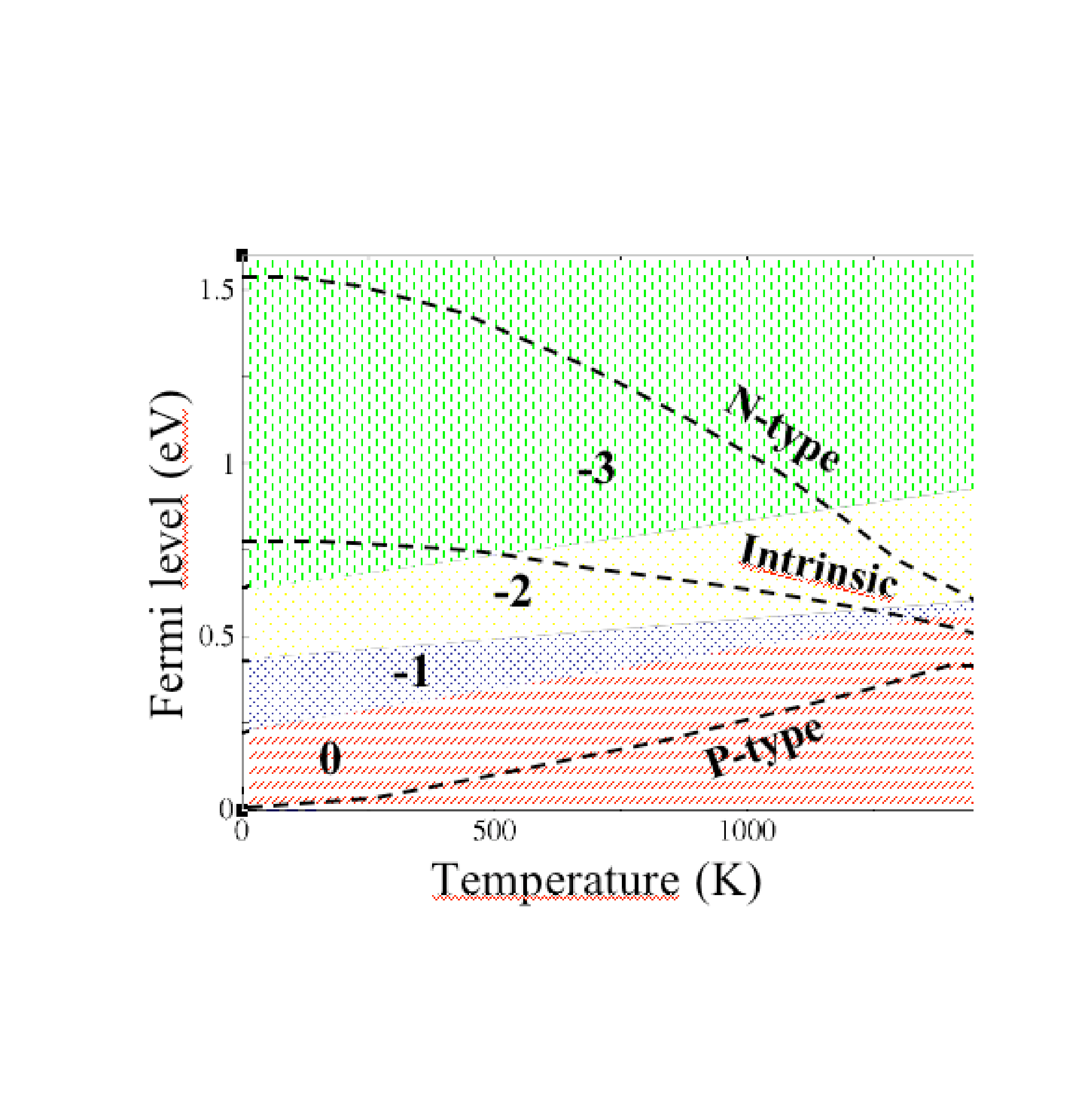}}}
\caption[Stability domains of $V_{Ga}^q$ in GaAs as function of
  temperature, indicated by the charge state.]{ Stability domains of $V_{Ga}^q$ in GaAs as function of
  temperature, indicated by the charge state. Each region denotes the
  temperature and the doping range where a charge state dominates self
  diffusion. The dashed lines show the typical position of the chemical
  potential as function of temperature for N, P and intrinsic conditions. At
  temperatures above 1000K the competition between different charges becomes
  significant.}
\label{fig:domains}
\end{figure}

These entropic effects are directly associated with changes in the phonon
density of states (DOS) as electrons are added to the system, modifying the
bonding strength in the region surrounding the defect. In the case of
$V_{Ga}$, the phonon DOS for the neutral vacancy shows the highest deviation
from the crystalline value. The effective impact of the vacancy on the lattice
is equivalent to the application of a local negative pressure: the transverse
acoustic (TA) modes move from the crystalline value by 3~cm$^{-1}$ to higher
frequencies (mode hardening), while the LA, LO, TO modes move to lower
frequency (mode softening) by 2~cm$^{-1}$. As the defect is progressively
charged, the DOS tends to recover the frequencies of the crystal DOS: TA modes
move to lower frequencies while LA, LO and TO move to higher frequencies
leading to an overall decrease in the entropy with the charge state.

Structurally, this behaviour is associated with the relaxation of the arsenic
atoms surrounding the empty site. In an unrelaxed model, As atoms form a
tetrahedron around the empty site with an interatomic distance of 3.95 \AA. As
electron are added to the relaxed system, the symmetry is conserved but volume
of the vacancy decreases. A detailed structural analysis reveals that the
addition of the first electron does not affect considerably the bond length
around $V_{Ga}$. As-As bonds decrease progressively to 3.37\AA\ for 0 and $-$1
charges and 3.36\AA\ and 3.35\AA\ for $-$2 and $-$3 charge states. To
compensate, since the simulation is performed at fixed volume, the Ga-As
bonds, for the atoms surrounding the vacancy, stretch out with the charge;
this effect propagates in the direction of the link until the borders of the
supercell especially on the high symmetry axes. The pressure increases locally
around the vacancy but it is released farthest away from it. The phonon
density of states suffers from a strong competition between the compression
and the expansion of bonds, and the electron-phonon interaction.

Finally, it appears that the DOS is under the effects of a small progressive
compression as electrons are added (TA softening and LA, LO, TO hardening)
that clearly overcomes the proposed phonon mode softening induced by
electron-phonon interaction studied earlier for GaAs in Ref.~\cite{4Hei76}.
Instead, our data show a brutal decrease of the phonon vibrational modes as a
neutral vacancy is introduced in a perfect crystal and a progressive
strengthening as electrons are added to the defect, agreeing with the
experimental observation~\cite{4Ish95} that LO($\Gamma$) phonon peak is shifted
toward the low frequency side (mode softening) and broadens its width
depending on the concentration of vacancies.

Results for our calculated formation entropies and enthalpies are summarized
in Table~\ref{tab:comparison}. Formation entropy for the neutral state, the
only one available in the literature, compare well with the {\it ab-initio}
values using anharmonic approximation by Bockstedte and Scheffler~\cite{4Boc96}
and a recent fit of experimental data by Gebauer {\it et al.}~\cite{4Geb03}
(further comparisons with earlier works can be found in these two references).
The formation enthalpy we calculated in a previous work~\cite{4Elm05} is in
good agreement with recent LDA calculation by Zollo {\it et al.}~\cite{4Zol04}
using 64 and 216 supercells, differences do not exceed 0.2 eV for $V_{Ga}^0$. We relaxed
$V_{Ga}$ in different charge states  using
spin-polarized LDA calculation but no Jahn-Teller distortion has been
observed. This confirms that the symmetry conservation of $V_{Ga}^q$ of is not
a drawback of LDA but a behaviour proper to cation vacancies~\cite{4Cha03}. To
our knowledge no calculation using GGA has been performed on this defect at this point.

\vspace{.5cm}
 \begin{table}
   \caption{Calculated formation entropies $ S_f$ and formation
   enthalpies $ H_f$ for $V_{Ga}^{0,-1,-2,-3}$ in GaAs are compared to
   earlier calculation and experimental fits (see the text)
 }
\label{tab:comparison}
\begin{ruledtabular}
\begin{tabular} {lcccc}

&Neutral &$-$1 & $-$2 &$-$3 \\
\\
\hline
\\
{\bf $ S_f$/$k_B$}   &$7.23^a$, $7.3^b$, $9.6\pm1^c$  & 4.46 &3.04 &0.755 \\
\\
 {\bf $ H_f$ (eV)} &$3.2^a$, $2.8^b$,$3.2\pm0.5^c$   &3.35 &3.78  &4.423\\ 
 \\

\end{tabular}
\end{ruledtabular}

$^a$ Present work\\ 
$^b${\it Ab-initio} calculation by Bockstedte and Scheffler~\cite{4Boc96}\\
$^c$ Fit of  concentration profiles   by Gebauer {\it et al.}~\cite{4Geb03}\\
\end{table}

Combining the results presented above, we obtain the free energies difference
between the various charge states: $F^{0/-1}$~=~0.22~+~2.74$k_BT$, $F^{-1/-2}$~=~0.42~+~1.42$k_BT$ and $F^{-2/-3}$~=~0.63~+~2.28$k_BT$. While the formation
entropies are known to suffer from finite-size effects, entropy differences
should be much more reliable. Formation entropy errors due to the
finite-size of the supercell are expected to be of similar magnitude for all
charge states and thus cancel out when only differences are considered
~\cite{4Raul04}. Contrary to earlier assumptions~\cite{4Van76}, the actual
temperature dependence of the ionization energies favors the formation of less
charged point defects at high temperature. Figure~\ref{fig:domains} shows the
stability domains within the experimental band gap for the dominant charge
states as function of temperature.

The transition from a charge state to another takes place as soon as the Fermi
level (under certain doping conditions) crosses the ionization energy line,
falling into a new stability domain. For P-type, the 0/$-$1 level is always
higher than the extrinsic Fermi energy suggesting that the neutral vacancy
remains stable at all temperatures. For intrinsic and N-type conditions,
however, the charge state of the vacancy is lowered as the temperature
increases. The critical temperature associated with these changes is located
in the range of 520-1500K where $V_{Ga}^{-2}$, $V_{Ga}^{-1}$ and $V_{Ga}^{0}$ start to
compete with $V_{Ga}^{-3}$.

\subsection{Uncertainties on the calculated Gibbs free energy}

The formation enthalpies and transition levels calculated with LDA involve
uncertainties that can exceed the vibrational effects, but that are more
likely to cancel out. Thus, one must be aware of possible source of errors
affecting the Gibbs free energy of formation. Formation entropies suffers from
errors associated with the calculation of vibrational density of
states~\cite{4Mat04}. For example, the microscopic structure of the defect
itself, which is possibly not obtained accurately. This may be a result of the
defect-defect interaction of the supercells, but also errors from the
construction of the pseudopotentials or the LDA may contribute. In general,
these errors only affect slightly the total energy calculation but are more
significant for vibrational modes. Moreover, uncertainty in the lattice
constant which causes the defect to exerts pressure on its environment and can
modify the equilibrium lattice constant of the defective supercell compared to
the LDA bulk value. Since we are only interested in Gibbs free energy differences we may expect a significant  amount of error cancellation. Recent tests on phonon frequencies of silicon vacancy in bulk silicon~\cite{4Alm03} showed that a cancellation of errors does tend to occur because  free energy differences  depend only weakly on the choice of lattice constant.

Uncertainties on formation enthalpy can come mostly from two sources. First,
the LDA band gap problem,  since it is well-known that LDA usually
underestimates the band gap of materials. The simplest way, used here, for
dealing with the band gap problem ~\cite{4Elm05} is to shift the conduction
band states up uniformly by the amount needed to reproduce the experimental
gap~\cite{4Bar84} (align the conduction band maximum (CBM) with the
experimental value, at 1.52 eV from the valence band maximum (VBM)). Since the
LDA can produce similarly large errors in the energies of the deep defect
states, it is also important to correct these errors. Therefore, defect states
with predominantly conduction band character experience the same upward shift
as the the conduction band states themselves, while leaving the defect states
with predominantly valence band character fixed relative to the valence band
edge. In our case, $V_{Ga}^q$ ionization levels are calculated by total-energy
energy differences, the correction we use here leaves these ionization levels
unchanged since they have a valence band character. Many other correction
schemes can be used with good efficiency as discussed in the work of Castleton
and coworkers~\cite{4Cas06}, we discuss below the impact of such corrections on
the fitting procedure. With current computational power limitations, a full GW
calculation, which would correct LDA errors, is unfortunately not possible for
the large supercells needed for defect studies. Therefore, LDA calculation is
the best choice currently since it provides a good estimate of the electronic
ionisation levels that enables us to fit efficiently experimental data.
Second, supercell corrections could also introduce additional errors on the
calculated formation enthalpy. In this work, total energies of charged defect
are corrected using the first order Makov-Payne method~\cite{4Mak96}. Many
articles~\cite{4Cas06, 4Shi05, 4Ger03} have recently shown that Makov and Payne
correction can be overestimated for highly charged defects.

\subsection{Fitting experimental measurements}

\begin{figure}[t]
\centerline{\rotatebox{-90}{\includegraphics[width=7cm]{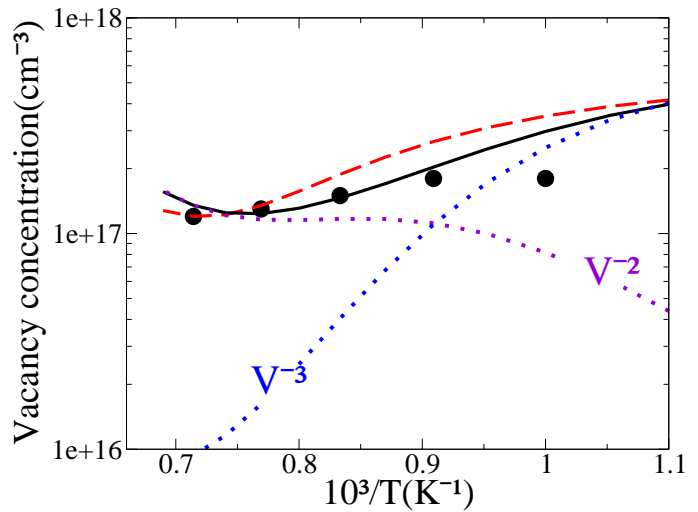}}}
\caption[Experimental measurement (filled circle), and fit
  (dashed line) for temperature dependence of the vacancy concentration, as
  reported by Gebauer {\it et al.}~\cite{4Geb03}.]{ Experimental measurement (filled circle), and fit
  (dashed line) for temperature dependence of the vacancy concentration, as
  reported by Gebauer {\it et al.}~\cite{4Geb03} .  Our calculation including
  $-$2 and $-$3 contributions (solid line) is shown together as well as
  separately (dotted line).}
\label{fig:Geb}
\end{figure}

\begin{figure}[t]
\centerline{\includegraphics[width=7cm]{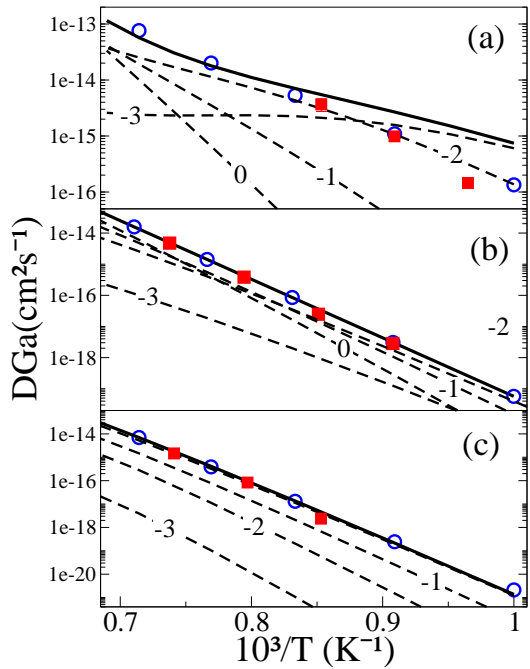}}
\caption{ Gallium self diffusion constant versus inverse
temperature for (a) N-type (b) intrinsic (c) P-type. Theoretical results for
$D^q_{Ga}$ (dashed lines) show the contribution of each charge state. The
transition temperature lie at the crossing of the curves. The total $D_{Ga}$
(solid) is in good agreement with the measured interdiffusion data (square)
and fit by Bracht {\it et al.}~\cite{4Bra99_2}(circle).}
\label{fig:DGa}
\end{figure}

We revisit the analysis of two apparently conflicting results by looking at
diffusion data of Bracht {\it et al.} and concentration profiles of Gebauer
{\it et al.} We first reproduce exactly the fits of Gebauer {\it et al.} and
Bracht {\it et al.} using exactly the same conditions described in the
respective original papers (dash lines in Fig.~\ref{fig:Geb} and squares in Fig.~\ref{fig:DGa}) and explained above. We then use the Gibbs free energy of
formation and all resulting ionisation levels calculated by LDA to generate
our fits. The contributions coming from each charge state is shown and  the total concentration and diffusion coefficient are plotted with
solid lines in Figs.~\ref{fig:Geb} and~\ref{fig:DGa}.

Using our calculated values, with only one free parameter to adjust the
prefactor ($C^*$ in equation~\ref{eq:bra2}), and $-$2 and $-$3 charges, we reproduce the experimental data of
Gebauer {\it et al.} with a good accuracy (Fig.~\ref{fig:Geb}) showing that
the $-$2 charge state is the main contributor to self-diffusion at high T.

Self-diffusion coefficient measured by of Bracht {\it et al.} are also
efficiently fitted. Comparing our calculated Ga self-diffusion parameters with
experimental data and the corresponding fits~\cite{4Bra99_2}, with an
adjustable prefactor $C^*$ we notice (see Fig.~\ref{fig:DGa}(b)) a strong
competition between $-$2, $-$1 and 0 charge states for intrinsic GaAs while
$-$3 does not contribute to the total concentration. Moreover, a progressive
charge reversibility occurs: $V_{Ga} ^{-2}$ dominates at high temperature
until $T^{(-2/-1)}_I$~=~1230K, then $V_{Ga} ^{-1}$ dominates until
$T^{(-1/0)}_I$~=~1345K. Even if details about the location of transition
temperatures are different, the total diffusion coefficient is very well
reproduced with our model too. Contrary to what was proposed by Tan {\it et
al.}~\cite{4Tan93}, our results support the experiment, indicating that under
intrinsic conditions $V_{Ga}^{-3}$ plays very little role.

For Si-doped GaAs, while Bracht {\it et al.} find that the concentration
decreases with temperature for $V_{Ga} ^{-2}$ but shows no contribution from
the $-$3 charge states, we see a more complex situation
(Fig.~\ref{fig:DGa}(a)): self-diffusion is mediated by $V_{Ga} ^{-3}$ at
temperature below $T^{(-3/-2)}_N$~=~1150K, while above this transition
temperature $-$2 charge state is dominant until $T^{(-2/-1)}_N$~=~1450K.
Finally, for P-type GaAs (Fig.~\ref{fig:DGa}(c)), we obtain results that are
very similar to Bracht: the neutral vacancy is dominant over the whole
temperature range.

As is seen in Figs.~\ref{fig:Geb} and~\ref{fig:DGa}, our model reproduces very
well all data at high temperature for different doping condition. The
agreement is weaker for N-type GaAs below 1100K, however. In this regime both
experimental data sets lie below our predictions. This is likely due to the
formation of vacancy complexes and slow Ga-self-diffusion which appear to
delay the establishement of equilibrium condition, as noted by Gebauer {\it et
al.}~\cite{4Geb03}. Thus, the measured concentration is a lower limit of the
calculated equilibrium concentration.

A gap or supercell correction that applies the same rigid shift to the
ionization levels would not change the observed sequential charge
reversibility, caused by entropic effects, but only the location of transition
temperature. In this particular case, if the ionization levels are shifted
upwards, the contribution of $-2$ charge state increases. Thus, the remaining
discrepancies that appears between experiment and theory visible in
Fig.~\ref{fig:DGa}(a) for the low temperature limit can be significantly
attenuated. At a first insight, we propose to use the simplest LDA
corrections, if entropy effects alone give unsatisfactory results, one can go
beyond by using more advanced gap correction schemes.

\section{Conclusion} 
\label{sec:conclusion} 

The present work shows that the deformation potential and the electron-phonon
interaction induced by charged vacancies affect the phonon spectrum in a much
more complicated way than was thought previously~\cite{4Boc96, 4Van76}.
Consequently, the picture about ionization levels obtained at 0K~\cite{4Elm05}
is not applicable in the experimental range as ionization free energies are
temperature dependent. We have found that the entropy of formation calculated
with LDA contributes to lower the free energy of formation at high temperature
when associated with empirical thermal effects on the Fermi level. Thus, the
formation of less charged Ga vacancies mediating self-diffusion in GaAs
becomes much more favored at high temperature. In spite of LDA limitations,
the resulting coherent Gibbs free energies of formation are a possible
solution to the apparently contradictory experimental results about the
dominant charge state for vacancy diffusion in GaAs. In addition to resolving
this controversy, our results underline the need to include thermal effects on
the Fermi level together with ionization entropies and ionization enthalpies.
Even though the Fermi level effect dominates the correction term, the smaller
contribution coming from ionization entropy is crucial for a correct
interpretation of experimental data.

{\it Acknowledgements.} We are thankful to Dr Hartmut Bracht for sending us his
data and a number of discussions. NM acknowledges partial support from FQRNT
(Qu\'ebec), NSERC (Canada) and the Canada Research Chair program. We thank the
RQCHP for generous allocation of computer time.

\end{document}